\documentstyle[emulateapj]{article}
\newcommand{\vv}{\vec{v}}
\newcommand{\vn}{\vec{n}}
\newcommand{\vf}{\vec{F}}
\newcommand{\mint}{\int\!\!\!\!\int\!\!\!\!\int}

\slugcomment{Astronomical Journal, 118, October 1999}
\lefthead{WANG}
\righthead{Contact discontinuities in Contact Binaries}
\begin{document}

\title{Contact Discontinuities in Models of Contact Binaries\\
Undergoing Thermal Relaxation Oscillations}

\author{Jian-Min Wang}
\affil{Laboratory of Cosmic-ray and High Energy Astrophysics,
Institute of High Energy Physics, CAS, Beijing 100039, 
and Chinese Academy of Sciences -- Peking University Joint 
Beijing Astrophysical Center (CAS-PKU.BAC), Beijing 100871, P.R. China, 
E-mail: wangjm@astrosv1.ihep.ac.cn}

\received{29 April 1999}
\accepted{10 June 1999}

\begin{abstract}
In this paper we pursue the suggestion by Shu, Lubow \& Anderson (1979) and 
Wang (1995) that contact discontinuity (DSC) may exist in the secondary 
in the expansion TRO (thermal relaxation oscillation) state. It is
demonstrated that
there is a mass exchange instability in some range of mass ratio for
the two components. We show that the assumption of {\it constant} 
volume of the secondary should be relaxed in DSC model. For {\it all} mass 
ratio the secondary alway satisfies the condition that no mass flow returns 
to the primary through the inner Lagrangian point. 
The secondary will expand in order to equilibrate the interaction 
between the common convective envelope and the secondary.
The contact discontinuity in contact binary undergoing thermal relaxation 
does not violate the second law of thermodynamics. The maintaining
condition of contact discontinuity is derived in the time-dependent model.
It is desired to improve the TRO model with the advanced 
contact discontinuity layer in future detailed calculations.


\end{abstract}


\section{Introduction}
W Ursae Majoris (W UMa) binary stars were first thought to be a particular
binary population due to their abnormal mass-radius relationship,
namely, the so-called Kuiper's paradox, $R_2/R_1=\left(M_2/M_1\right)^{0.46}$ 
(Kuiper 1941).  These particular binaries appear
to consist of two main-sequence stars that possess photospheres exhibiting
the almost same effective temperatures for the two components despite the
fact that typical mass ratio in a system is 0.5. 
It was originally proposed that a common convective
envelope may be formed due to dynamic equilibrium (Osaki 1965), and
mass and energy transfers would take place in CCE in order to interpret
the Kuiper's paradox (Lucy 1968) although the specific
mechanism of energy transfer for the circulation has not been fully
understood (Robertson 1980, Sinjab, Robertson \& Smith 1990, Tassoul 1992). 
It is now firmly believed that W UMa stars are contact binaries in which
both components are full of their Roche lobe, showing strong interactions
(Mochnacki 1981). Lucy's iso-entropy model (1968) as a zero-order model with 
thermal equilibrium, however, cannot explain
the color -- period diagram by Eggen (1967) which leads to the 
establishments of two parallel first-order theories of thermal relaxation
oscillation (TRO) and contact discontinuity (DSC). TRO 
model was advanced by Lucy (1976), Flannery (1976), and Robertson \& Eggleton
(1977), who suggest that contact system can not reach thermal equilibrium
at dynamical equilibrium configuration and may thus undergo thermal
relaxation oscillation. DSC theory was proposed by Shu and his
collaborators [Shu, Lubow \& Anderson 1976, 1979, Lubow \& Shu 1977, also
see Biermann \& Thomas (1972) and Vilhu (1973) for some earlier elements
of DSC model] who hypothesize contact binary can attain thermal and dynamical 
equilibrium but there is a temperature inversion layer in the secondary.
With great attempts the so-called Kuiper's paradox and period - colour diagram
may be resolved by the two different hypotheses 
independently. However, both theories have some difficulties to explain 
observations, such as the so-called W-phenomenon, i.e., the less
star is hotter than the massive component (Binnendijik 1970)
(see a concise summary of observations and theory by Smith 1984).
Especially there are great debases between the two in nature in their simplest 
version (K\"ahler 1989). Observational studies continue and the 
theoretical controversy still remains (Ruci\'nski 1997). These imply that the 
first-order theories of contact binary (i.e. TRO and DSC) should be improved.

The intensive disputes by the two contending schools (Lucy \& Wilson 
1979, Shu, Lubow \& Anderson 1979) lead to an intriguing suggestion by 
Shu, Lubow \& Anderson (1979), Shu (1980) (from theoretical viewpoint) 
and Wang (1995) (from the analysis of observational data) that TRO 
theory needs the contact discontinuity in some phases. Some observations 
seem to support TRO theory (Lucy \& Wilson 1979, Hilditch et al 1989, 
Samec et al 1998). Although some criticisms about DSC model exist (Shu, Lubow 
\& Anderson 1980), this theory is still attractive because it is successful 
in many aspects (Smith 1984). The contact discontinuity may be ironed out 
within the thermal timescale (Webbink 1977, Hazlehurst \& Resfdal 1978, 
Papaloizou \& Pringle 1979, Smith, Robertson \& Smith 1980) 
in a steady state, however, the existence of time-dependent contact 
discontinuity can not be excluded because it does not violate the second 
law of thermodynamics (Papaloizou \& Pringle 1979).  However a detailed
analysis is needed for this. It is highly desired to reconcile the two 
theories not only for removing the discrepancies but also for explaining
more detailed observations (Shu 1980).

The difficulties of pure TRO and DSC models in explanation of W-phenomenon
motivate us to explore the possibility to develop a second-order theory.
The interaction between the secondary
and the common convective envelope is thought as an important role in the
W-phenomenon, Wang (1994) find that the W-phenomenon can be explained
by the released gravitational energy of the secondary through its contraction 
corresponding to the TRO contracting phase in W-type contact systems. This is 
encouraging and leads to the suggestions by Wang (1995) from a sample with 32 
contact systems that the A-type systems may undergo thermal relaxation 
oscillation with contact discontinuity whereas the contraction of secondary 
in W-type systems irons out this contact discontinuity.

The over-riding virtue of a contact discontinuity is that it gives a
clear mechanism for making the secondary physically larger, and the
primary physically smaller, than their main-sequence single-star
counterparts, as is needed to satisfy the Roche-lobe filling requirements
of the Kuiper paradox (Lubow \& Shu 1977). On the other hand, if the
system cannot be maintained in steady state by heat-carrying flows,
then the capping of the radiative heat flow from the secondary by the 
hotter overlying common envelope should lead to an expansion of the
secondary, with a resulting transfer of mass from the secondary to the
primary. Such Roche-lobe overflow, from a less massive star to a more
massive star to a more massive one, is known to occur slowly, so the
ultimate breaking of contact caused by the expansion of the binary orbit
takes a relatively long time. Once contact has been broken, however, the common
envelope disappears; the secondary is no longer capped; and it will begin
to shrink toward its normal single-star size. Conversely, because the
larger area of the common envelope is no longer available to carry away
much of the primary's interior luminosity, the primary can no longer be
maintained at its suppressed contact size, and it will begin to overflow
its Roche lobe. The transfer of mass from a more massive star to a less
massive one is known to be unstable (e.g. Paczy\'nski 1971), and the rapid
shrinkage of the binary orbit causes the system to come into contact again.
The re-establishment of the common envelope and the capping of the secondary
results in its refilling its Roche lobe. Thus, would the DSC hypothesis
provide the physical mechanism for the TRO hypothesis, together with a
justification why the duty cycle is long for the contact phase and short
for the semidetached phase, as is required by the observational statistics.
The rest of this paper attempts to establish the above ideas on a more
quantitative basis.

This paper is organized as following: the instability of Roche lobe and its 
operation in contact system are found in Sec.2; the surviving
condition of DSC layer is derived from  the thermodynamics in Sec. 3; and 
the conclusions are remarked in last sectionr.

\section{The Roche lobe instability and DSC model}
It is generally believed that the two components of W UMa stars  
share an optically thick common convective
envelope due to the dynamics equilibrium
(Osaki 1965, Mochnacki 1981). The redistribution of the total luminosities
(the plus of luminosities of each star), which takes place in CCE,
deals with comprehensive fluid processes (Lucy 1968). Why and how to 
redistribute the luminosties is the main task to theoreticians. 
The debate of the exsiting theories of contact binaries had been attracted 
much consideration between 1970s and 1980s (Lucy \& Wilson 1979, Shu 1980, 
Shu, Lubow \& Anderson 1981). Shu (1980) clearly stated that  the two 
superficially distinct theories are complementary
with the crucial theoretical issue to be resolved being the secular
stability of temperature inversion layer from his thought-provoking analysis. 
Here we argue that DSC layer is a natural results of TRO theory via
the mechanism of Roche lobe instability,
showing the presence of DSC layer during the expansion TRO phase.

In the following discussions we assume that the total mass and angular momentum
are conserved, neglecting the spin angular momentum of two components. These 
assumptions are basic and the same in TRO theory, but they are unnecessary 
in the DSC 
theory. In principle, the two assumptions put more strong constraints on the 
theoretical model.  In the conserved systems there are mainly two other 
parameters: mass ratio $q$, and mass ratio changing rate due to mass exchange
$\dot{q}$, to determine 
the structure of the contact binaries in TRO theory. 
The most serious shortcoming of TRO (mentioned in the previous section)
is a strong indicator that we should relax some of assumptions in
TRO model. One possible way to remove this shortcoming is to supplement 
the interaction between CCE and the component. This inclusion may reconcile 
the two contending schools each other (Wang 1995). 

We first show that the instability of mass exchange may prevent from the 
mass in the secondary being pushed into the primary through the inner 
Lagrangian point $L_1$ due to the lid effects of 
CCE placed on the secondary (Shu, Lubow \& Anderson 1976).  For a contact 
system with total angular momentum ${\cal J}$ in a circular orbit and total 
mass ${\cal M}=M_1+M_2$, the separation between components reads
\begin{equation}
{\cal D}=\left(\frac{{\cal J}^2}{G{\cal M}^3}\right) \frac{(1+q)^4}{q^2},
\end{equation}
where $q=M_2/M_1$ (for the convenience we take $q\leq 1$), and $G$ is 
the gravitation constant. The Roche lobe 
radius $R_{\rm L}$ of the secondary approximates for all mass ratio
(Eggleton 1983)
\begin{equation}
r_{\rm L}=\frac{R_{\rm L}}{{\cal D}}
         =\frac{0.49}{0.6+q^{\frac{2}{3}}\ln(1+q^{-\frac{1}{3}})}.
\end{equation}
The Roche lobe of the primary will be obtained when we replace $q$
by $1/q$.
It is important to note that the Roche lobe is changing due to the mass 
transfer between the two components. The variation rate of the Roche lobe 
due to mass transfer between the two components reads
\begin{equation}
\frac{d\ln R_{\rm L}}{dq}=\frac{2r_{\rm L}}{3q^{\frac{1}{3}}}
       \left[\frac{1}{1+q^{\frac{1}{3}}}-2\ln (1+q^{-\frac{1}{3}})\right]+
       \frac{2(q-1)}{q(1+q)},
\end{equation}
and then we have the timescale for this change with the helps of
$d\ln R_{\rm L}/dt=(d\ln R_{\rm L}/dq)(dq/dt)$
\begin{equation}
t_{\rm R_{L}}=\left(\frac{d\ln R_{\rm L}}{dt}\right)^{-1}
             =f(q)t_{\rm M},
\end{equation}
where $t_{\rm M}$ is the timescale of mass transfer defined as
\begin{equation}
t_{\rm M}= \frac{M_1}{\dot{M}_1},
\end{equation}
here the parameter $\dot{M}_1$ is the rate of mass transfer, and the function 
$f(q)$ is 
\begin{equation}
f(q)=\left\{ \frac{2(1-q)}{q}-\frac{2r_{\rm L}}{3q^{\frac{1}{3}}}
       \left[\frac{1}{1+q^{\frac{1}{3}}}-2\ln (1+q^{-\frac{1}{3}})\right]
       (1+q)\right\}^{-1}.
\end{equation}
The function $f(q)$ represents the ratio of the two 
timescales. We have calculated the function $f(q)$ in Figure 1, showing 
its value for the range of $q$ from 0.0 to 1.0. 
If $f(q)>0$ then Roche 
lobe will expand with increases of $q$ or shrink with decreases of
$q$. If $f(q)<0$ then Roche lobe will shrink with the increases of 
$q$ or expand with decreases of $q$. It is very important to 
address that if $|f(q)|<1$ then the expansion or shrinkage will be rapid 
than the process of mass transfer in the contact system known from equation
(4). From Figure 1 the Roche lobe of the secondary expands with the 
increases of mass ratio. The expansion timescale is shorter than that of
gaining mass from the primary until the mass ratio $q>0.8$,
indicating that  the secondary is
capable of swallowing more mass when $q<0.8$. In this range of mass ratio
the mass gaining is unstable. The Roche lobe of the primary
will shrink due to the mass exchange. There also is an instability of
mass exchange in the range of $q<0.35$. This implies that the timescale
of Roche lobe shrinkage due to mass transfer is shorter than that of
mass transfer. This means that Roche lobe shrinks more rapid than the
mass loss. This mass exchange instability plays important role in the
structure of contact binaries. The maximum of mass ratio for the instability
of the primary is 0.35. It is believed that the mass transfer will be
more efficient when $q<0.35$ with the presence of the instability of mass
exchange. Thus it is expected that the mean mass ratio
of A-type systems will be less than that of W-types. This is consistent
with the observations. It should be noted that here we do not specify the 
mechanism for the energy and mass transfer. 
Of course the direction of mass exchange between the two components
is determined by the relative potential of the star surface. Here we are trying
to show the instability of mass exchange, namely, represented by $|f(q)|<1$.
 
In the original DSC version the rising convective elements interior to the
Roche lobe of the secondary cannot penetrate into the common convective
envelope because the resulting buoyancy deficit opposes such penetration.
SLA76 argued that there is a mass flow pushed by the slight excess of 
pressure due to a slight heating of the interior of the secondary under 
{\it constant} volume. It is very important to note that this mass flow 
process is based on the assumption with {\it constant} volume of the 
secondary. The mass flow was estimated by SLA79, however, their estimation 
follows up another assumption that {\it all} the energy transferred from 
the primary to the secondary radiates again from the secondary, neglecting 
the interaction between CCE and the secondary. 
Now we can work out a condition inhibits the returning of mass flow to
the primary through the inner Lagrangian point $L_1$. If there is no excess
pressure, then the mass flow stops. This is equivalent to
$d\bar{\rho}/dt\leq 0$ if the {\it contact discontinuity} being lower than 
the temperature of CCE survives, namely, $T=$constant (we assume the gas is
ideal), where $\bar{\rho}$ is the mean mass density within the Roche lobe
of the secondary defined as $\bar{\rho} \propto M_2/R_{\rm L}^3$.
We thus  have
\begin{equation}
\frac{d\ln \bar{\rho}}{dt}\approx \frac{d}{dt}\left[\ln \left(
                          \frac{q}{1+q}\right)\right]-\frac{3}{t_{R_{\rm L}}},
\end{equation}
then the condition no returning of mass flow reads
\begin{equation}
t_{\rm R_L}\leq 3qt_{\rm M}.
\end{equation}
We draw the line $t_{\rm R_L}=3qt_{\rm M}$ in Figure 1. It is obvious that
all the value of $f(q)$ is always less than $3q$. This means that all the
cases satisfy the condition that no mass flow returns to the primary even
beyond the mass exchange instability. Therefore the assumption of 
{\it constant} volume of the secondary should be relaxed in the
advanced DSC model. The contact discontinuity is time-dependent from
this viewpoint at least, coinciding with that DSC layer could be
maintained in a time-dependent model (Papaloizou \& Pringle 1979).

\section{Thermodynamics of DSC layer}
By defining the thermal timescale as 
$t_{\rm Th}=\int_{M-\delta m}^{M}(4\pi r^2\rho v_c)^{-1}dm$, Webbink(1977)
first showed that the thermal diffusion time scale in the common envelope is
typically of the same order as the dynamical timescale (is roughly of
one orbital period). This makes the contact discontinuity disappear 
within one orbital period. We call the thermal diffusive process
as interaction $\epsilon$.
Papaloizou \& Pringle (1979) show the steady contact discontinuity violates 
the second law of thermodynamics, but the time-dependent contact 
discontinuity may exist. However in the time-dependent
model it is the interaction $\epsilon$ that keeps the contact discontinuity
in contact system undergoing thermal relaxation oscillation. 
The controversy of inner
structure may be removed by this kind of interaction (Wang 1995).  
With the helps of the conservation of mass and momentum 
we can rewrite the energy equation beyond the energy generation region as
\begin{equation}
\rho \frac{\partial}{\partial t} (\Psi+Ts)=-\rho T\vv \cdot \nabla s
     -\nabla \cdot \vf +\epsilon,
\end{equation}
for the inviscid fluid (e.g. Webbink 1977, and 1992 ApJ, 396, p378 for the 
erratum), where $t$ is time; $\rho$, the density; $T$, the temperature; 
$\vv$, the velocity; $s$, the specific entropy; $\vf$, the energy flux 
radiated from the star; $\epsilon$, the energy density absorbed by the 
secondary in the unit time due to the interaction with CCE; and $\Psi$, 
the gravitational energy per unit mass. Following the assumption by Shu, 
Lubow \& Anderson (1980) that the specific entropy $s$ can be decomposed 
in terms of a barotropic and a baroclinic one as 
$s=s_0(\Psi_{\rm D})+s_1(\vec{x},t)$, we integrate the above equation over 
the volume enclosed by the equipotential surfaces $C$ and $D$, and obtain 
\begin{eqnarray}
\frac{dS}{dt}&=& s_0(\Psi_D)\frac{d(\Delta M)}{dt}-
              \mint_{\rm CD}\frac{\rho}{T} 
              \frac{\partial \Psi}{\partial t}dV+
              \frac{\epsilon}{T}\Delta V \nonumber\\ 
          &   & -\oint_{\rm CD}\rho s_1(\vec{x},t)\vv \cdot \vn dA-
              \mint_{\rm CD}\frac{1}{T} \nabla \cdot \vf dV,
\end{eqnarray}
where $S=\mint \rho s dV$, $\Delta M=\mint \rho dV$, 
$\Delta V=\mint dV$ is the 
volume enclosed by the two surfaces $C$ and $D$, $dA$ is the area of the 
surface of contact discontinuity, and $\vn$ is its normal vector. For the 
time-dependent case we assume that the last two terms offset approximately as
in steady case (Shu, Lubow \& Anderson 1980), thus we have more physically
concise form of equation (10)
\begin{equation}
\frac{dS}{dt}=s_0(\Psi_D)\frac{d(\Delta M)}{dt}-
              \mint_{\rm CD}\frac{\rho}{T}
               \frac{\partial \Psi}{\partial t}dV+
              \frac{\epsilon}{T}\Delta V,
\end{equation}
The first term of right hand in equation (11) represents the entropy increases
due to mass exchange between CCE and the secondary, the last term does the
same meanings but due to energy exchange, the second term does the entropy
decreases of entropy due to the Roche lobe expansion. 
The enclosed volume is an open system undergoing mass and energy exchanges
with its surroundings rather than an isolated volume. This equation also tells 
us the resulting expansion due to the interaction $\epsilon$ if the contact
discontinuity layer survives: 1) if there is only exchanges of energy by 
thermal diffusion, namely, $\Delta M=$const, we have
\begin{equation}
\mint_{\rm CD}\rho \frac{\partial \Psi}{\partial t} dV \geq \epsilon \Delta V.
\end{equation}
This clearly states that the surviving of contact discontinuity must be 
provided by the expansion. Detailed calculation should be done in the future. 
2) There is a mass exchange between CCE and the secondary accompanying the
energy interaction, i.e., 
$d(\Delta M)/dt >0$, the expansion is at least 
\begin{equation}
\mint_{\rm CD}\rho \frac{\partial \Psi}{\partial t} dV \geq \epsilon \Delta V
             +s_0(\Psi_D)T\frac{d(\Delta M)}{dt}.
\end{equation}
This equation predicts the secular change of orbital period due to the
shift of mass ratio. 3) If the secondary keeps {\it constant} volume as 
originally suggested by Shu, Lubow \& Anderson (1976), the term 
$d\Psi/dt=0$, we always have $dS/dt\geq 0$ which means
the discontinuity will be ironed out with the thermal diffusive timescale.
The only possible way to relax this condition is the inclusion of the
changes of Roche lobe. This way will permits us unifying the two contending
hypotheses. According the simplest version of star structure, equations
(12) and (13) will provide the expansion velocity
\begin{equation}
v_{\rm int}\geq R_2\left(\frac{\epsilon}{\bar{\epsilon}}\right)
  \left(\frac{\Delta V}{V_2}\right)
  +Ts_0(\Psi_{\rm D})g_2^{-1}
   \frac{d(\ln \Delta M)}{dt},
\end{equation}
where ${\bar{\epsilon}}=(GM_2\Delta M/R_2)/V_2$ is the mean density of the 
gravitational energy between the secondary and CCE,
$g_2=GM_2/R_2^2$ the gravitational acceleration,
$V_2$ is the volume of the secondary. One should note that in the above
estimation we neglect the down directed propagation of energy to the interior
with dynamical timescale. 
Therefore the present estimation is somewhat higher than that of the actual.
Both of the mass exchange and energy interaction 
result in the expansion of the secondary, we thus have 
the minimum velocity
\begin{equation}
V= \max (v_{\rm int},v_{\rm R_L}),
\end{equation}
where $v_{\rm R_L}=dR_{\rm L}/dt$ is the expanding velocity of the Roche lobe.
This is the condition maintaining contact discontinuity. From the viewpoint 
of total energy (by the nuclear) conservation, the exhausted energy (i.e. 
$\epsilon$) to expand the secondary lowers the re-radiating efficiency
of the transferred energy from the primary. The lower temperature of the 
secondary than the primary may be an indicator of the presence of contact 
discontinuity. 

\section{Conclusions and Discussions}
Introducing a discontinuity of temperature by Shu and his collaborators 
(1976, 1979) the thermal instability of binary (Lucy 1976, 
Flannery 1976) can be suppressed, but its maintenance of DSC layer opens. 
In this paper we try to construct the physical scenario of time-dependent
model of contact binary. Only two assumptions that total mass and angular
momentum of the contact system are conserved are employed in this paper.
It is found that the mass exchange results in
the instability of Roche lobe in some ranges of mass ratio. We show
that this instability always satisfies the condition that keeps the
mean density of the secondary $d\ln \bar{\rho}/dt\leq 0$. Therefore
it ensures that no mass flow returns to the primary through the inner
Lagrangain point $L_1$.
The second-order theory predicts that the contact binary may be in
oscillations take place about a state with a contact discontinuity.
The temperature differences of DSC layer across the interface is
determined by the expansion velocity.

The existing TRO and DSC theories (Lucy 1976, Flannery 1976, Robertson
\& Eggleton 1977; Shu, Lubow \& Anderson 1976, 1979)
neglect the effects of interaction $\epsilon$ between CCE and the secondary.
Here we argue that it is the contact discontinuity layer that results in
the interaction between CCE and the secondary and in the meanwhile it is
the interaction that maintains the contact discontinuity.
We find this interaction $\epsilon$ can result in some interesting
issues. First it is the reason why the temperature of the secondary in A-types
is lower than that of the primary. Second the maintenance of contact 
discontinuity needs faster mass transfer which breaks down the  
deep contact. Thus the shortcoming of TRO theory will be removed.
It is highly desired that the unification of TRO theory and DSC model
should be calculated in order to discover the nature of the contact
binaries.

In the present work we do not specify the mechanism of thermal diffusion 
process. Although we have not performed the time-dependent model unifying
TRO and DSC hypotheses, this time-dependent unified
theory might give some predictions. First the secular behavior 
of period change of A-type systems is violent than that in W-type system 
in order to survive the existence of contact discontinuity.  Second, the 
maintenance of contact discontinuity may lead to the radial oscillation of 
the secondary with period from a few to several ten minus. The interaction 
between CCE and the secondary drives such a oscillation similar to the  
$\kappa$-mechanism working in other types of stars. It is thus 
expected to find the light variation during the primary eclipse as 
another probe of contact discontinuity in A-type systems.

{\acknowledgements
The author would like to express his honest thanks to the referee,
Professor Frank H. Shu, for his input of physical insight to the title and
the fourth paragraph of the first section, enhancing the scientific clarity
of this paper.  I appreciate 
the useful discussions with Drs. Zhanwen Han and Fangjun Lu.
This project is supported by Climbing Plan of The Ministry of Science and 
Technology of China 
and the Natural Science Foundation of China under Grant No. 19800302.}


\begin{figure}
\vspace{4.5cm}
\epsscale{1.0}
\plotfiddle{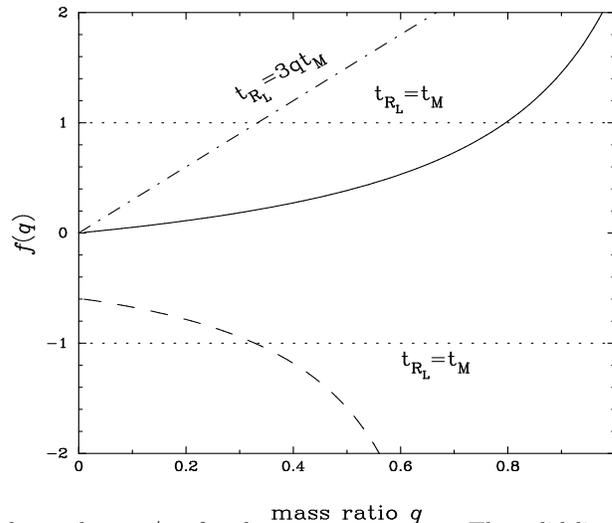}{90pt}{-0}{60}{55}{-160}{-50}
\vspace{-23mm}
\caption{\small The function $f(q)$ shows the $t_{\rm R_{L}}/t_{\rm M}$
for the two components. The solid line represents that of the secondary,
and the dashed line does the primary. See detail in the text.}

\label{fig1}
\end{figure}

\end{document}